\documentclass[10pt,jpa]{iopart}
\usepackage{iopams,bm,bbm,amssymb,amsbsy,graphicx,hyperref}

\usepackage[latin1]{inputenc}
\newcommand{\mathsym}[1]{{}}

\newcommand{\ket}[1]{|#1\rangle}
\newcommand{\bra}[1]{\langle#1|}

\newcommand{\eq}[1]{Eq.~(\ref{#1})}

\begin{document}
\title{Faithful nonclassicality indicators and extremal  quantum correlations in two-qubit states}
\author{Davide Girolami$^1$, Mauro Paternostro$^2$, and Gerardo Adesso$^1$}
\address{$^1$School of Mathematical Sciences, University of Nottingham, University Park, Nottingham NG7 2RD, United Kingdom\\
$^2$School of Mathematics and Physics, Queen's
University, Belfast BT7 1NN, United Kingdom}

\begin{abstract}
The state disturbance induced by locally measuring a quantum system yields a signature of nonclassical correlations beyond entanglement. Here we present a detailed study of such correlations  for two-qubit mixed states. To overcome the asymmetry of quantum discord and the  unfaithfulness of measurement-induced disturbance (severely overestimating  quantum correlations), we propose  an ameliorated measurement-induced disturbance  as nonclassicality indicator, optimized over joint local measurements, and  we derive its closed expression for relevant two-qubit states. We study its analytical relation with discord, and characterize the maximally quantum-correlated mixed states, that simultaneously extremize both quantifiers at given von Neumann entropy: among all two-qubit states, these states possess the most robust  quantum correlations against noise.
\end{abstract}


\date{April 28, 2011}


\bigskip

\section{Introduction}\label{secI}

The qualitative and quantitative study of genuinely quantum correlations in multipartite states is nowadays one of the most central and debated topics in quantum information science, involving both theoretical and experimental  lines of research \cite{piani,cinesi,DattaQC,OZ,HV,terhal,MID,wu,dissonance,pianietal,varissimo,operdiscord,discordsymm,james,maiorca}. Satisfactorily addressing when and to what extent a quantum system exhibits a departure from a purely classical behavior, and defining reliable signatures of its quantumness,  is crucial to achieve a deeper insight on the foundations of quantum mechanics, on the quantum-to-classical transition in complex systems and living organisms \cite{qbio}, and might open the way to novel technological developments in quantum engineering, communication and information processing \cite{qtech}.
While the quantum character of a pure state is conclusively revealed by a Bell inequality violation, or by its resource character for protocols such as teleportation, mixed  states entail a more involved yet exciting case where nonclassicality, nonlocality and entanglement do not necessarily coincide. Even highly mixed states with little or no entanglement can be useful for quantum computing, thus displaying strong signatures of quantumness in the form of nonclassical correlations beyond entanglement~\cite{DattaQC}. In an attempt to characterize such correlations and understand their role, several measures were proposed to pinpoint the various facets of nonclassicality~\cite{OZ,HV,terhal,MID,piani,wu,dissonance,pianietal}.
Among them, the increasingly popular {\it quantum discord}~\cite{OZ} and the more easily computable {\it measurement-induced disturbance}
(MID)~\cite{MID} have attracted considerable attention \cite{varissimo,operdiscord}.

Whilst the theory of entanglement has been amply developed, leading to a set of well motivated requirements that any {\it bona fide} entanglement measure should satisfy~\cite{horodecki}, a similar formal backbone is missing for general nonclassicality indicators.
This leaves room for  the drawing of physically incorrect conclusions about the nature of correlations in a quantum state, should
inappropriate measures be employed. To render the paradigm of quantum correlations beyond entanglement widely accessible and appealing experimentally, and be able to  reveal truly quantum rather than classical features in laboratory systems, one should first  identify those quantifiers of quantum correlations that are {\it faithful}, namely  vanishing on all {\it classical-classical} states \cite{piani} (classical probability distributions embedded in a density matrix), and avoid the use of unfaithful ones. Desirable requirements for a  quantum correlation measure would further be its symmetry with respect to swapping the subsystems (correlations should not depend on which party is probed) and its vanishing {\it only} on classical-classical states (strong faithfulness).

In this paper, motivated by these premises, we assess the characterization of quantum correlations focusing on the paradigmatic instance of (generally mixed)
two-qubit states. We first deploy a quantitative benchmarking test  of quantum discord and MID  as tools to investigate the interplay
between quantum correlations and global state mixedness. We find that the nonoptimized nature of MID \cite{MID,wu} makes such indicator unfaithful, being nonzero and even maximal on some classical-classical states. On the other hand,  due to its asymmetric definition \cite{OZ}, discord is not strongly faithful   \cite{discordsymm}, as it does not reliably detect the fine  discrimination between classical-classical and so-called {\it classical-quantum} states \cite{piani}, which still possess some quantum correlations, but exhibit zero discord.
We thus propose to employ in general an {\it ameliorated} version of MID as a measure of quantum correlations, which we refer to as ``AMID'', operatively associated to the minimal state disturbance upon joint local measurements (in the spirit of Refs.~\cite{terhal,piani,wu,discordsymm}).  AMID is symmetric by construction,  plays a clear role as quantum complement to the `classical mutual information' \cite{terhal}, provides a tight upper bound to discord,
and vanishes on  and only on all classical-classical states,  yielding a strongly faithful quantification of quantum correlations.
Notably, we provide an analytical recipe to evaluate AMID on general two-qubit states, which leads to closed formulas for some relevant families of states.

Reaching beyond the recent efforts of Refs.~\cite{james,maiorca}, and inspired by an analogous earlier study of maximally entangled mixed states (MEMS) \cite{munro}, we furthermore identify and characterize rigorously  the {\em maximally quantum-correlated mixed states} of two qubits at given values of the global von Neumann entropy. In the entropic plane, discord and AMID admit exactly the same (rather structured) set of universal extremal states, that collectively embody the maximal robustness of two-qubit quantum correlations against decoherence.
Our investigation sheds unforeseen light onto a topic of vast theoretical and practical interest.
The families of extremal states identified here can be experimentally engineered by means of light-atom interfaces~\cite{noiMegan} or all-optical setups, and the quantum correlations can be directly measured in laboratory via suitable local  detections or computed from  the tomographically reconstructed density matrices \cite{cinesi,mauroexp}.

The paper is organized as follows. In Sec.~\ref{secMQ} we define discord and MID and explore their distribution as a function of the global state mixedness (von Neumann entropy) for general two-qubit states. We identify the maximally quantum-correlated mixed states of two qubits, and highlight sharp unfaithful features of the MID measure. In Sec.~\ref{secAMID} we introduce the AMID and study its properties, providing analytical progresses on its evaluation for relevant two-qubit states. A comparison between the three nonclassicality indicators is presented and reveals how the AMID admits the same extremal states as discord at fixed global entropy. We summarize our results in Sec.~\ref{secC}.

\section{Maximally quantum-correlated mixed states of two qubits: Discord versus MID}\label{secMQ}

  Let us fist introduce two of the nonclassicality indicators studied in our paper. Ollivier and Zurek associate quantum discord to the discrepancy between two classically equivalent versions of mutual information. The latter is a widely accepted  measure of total correlations, defined, for a bipartite quantum state $\varrho_{AB}$, as ${\cal I}(\varrho_{AB}){=}{\cal S}(\varrho_A){+}{\cal S}(\varrho_B){-}{\cal S}(\varrho_{AB})$. Here, ${\cal S}(\varrho){=}{-}{\rm Tr}[\varrho\log_2\varrho]$ is the von Neumann entropy (vNE) of the arbitrary two-qubit state $\varrho$ and $\varrho_{j}$ is the reduced density matrix of party $j{=}A,B$. Alternatively, one can consider the expression ${\cal J}^\leftarrow(\varrho_{AB}){=}{\cal S}(\varrho_A){-}{\cal H}_{\{\hat\Pi_i\}}(A|B)$ (the one-way classical correlation \cite{HV}) with ${\cal H}_{\{\hat\Pi_i\}}(A|B){\equiv}\sum_{i}p_i{\cal S}(\varrho^i_{A|B})$ the quantum conditional entropy associated with the the post-measurement density matrix $\varrho^i_{A|B}{=}{\rm Tr}_B[\hat\Pi_i\varrho_{AB}]/p_i$ obtained upon performing the complete projective measurement $\{\Pi_i\}$ on system $B$ ($p_i{=}{\rm Tr}[\hat\Pi_i\varrho_{AB}]$). Quantum discord is thus defined as \begin{equation}\label{eqDisc}{\cal D}^\leftarrow = \inf_{\{\Pi_i\}}[{\cal I}(\varrho_{AB}) - {\cal J}^\leftarrow(\varrho_{AB})]\,,\end{equation} where the infimum is calculated over the set of projectors $\{\hat\Pi_i\}$ \cite{OZ}. Discord is, in general, asymmetric as ${\cal D}^\leftarrow{\neq}{\cal D}^\rightarrow$ with ${\cal D}^\rightarrow$ obtained by swapping the roles of A and B. No closed formulas are known  for ${\cal D}^\leftarrow$ on general two-qubit states, other than special cases \cite{luo,alber}.

Luo introduced MID starting from the observation that a bipartite state containing no quantum correlations commutes with the operators describing any complete projective measurement~\cite{MID}. On the other hand, although a state $\varrho_{AB}$ may be intrinsically nonclassical, any complete bi-local projective measurement makes it classical as a result of a decoherence-by-measurement process~\cite{MID}. MID is thus defined by restricting the attention to the complete projective measurement $\{\hat\Pi_{kl}{=}\hat\Pi_{A,k}\otimes\hat\Pi_{B,l}\}$ determined by the eigen-projectors $\hat\Pi_{j,k}$ of $\varrho_{j}$~($k{=}1,2$) and reads
\begin{equation}\label{eqMID}{\cal M}(\varrho_{AB}) = {\cal I}(\varrho_{AB}) - {\cal I}(\varrho^{\hat\Pi}_{AB})\,,\end{equation}
where $\varrho^{\hat\Pi}_{AB}$ is the state resulting from the application of $\hat\Pi_{AB}$.

 We begin our analysis by investigating quantum correlations versus global state mixedness (vNE) for two qubits, looking in particular for the families of extremal states \cite{notejamesvne}.  We have  generated up to $2{\times}10^6$ random   two-qubit density matrices, uniformly in the space of Hermitian, positive semidefinite matrices, and for each of them we have evaluated ${\cal S}$, ${\cal M}$ (analytically) and ${\cal D}^\leftarrow$ (numerically). Most notably, although our  study has been performed using unconstrained density matrices, we have found that the so-called $X$ states of the form
\begin{equation}
\label{Xstates}
\varrho^{X}_{AB}=\left(
\begin{array}{cccc}
\varrho_{11}&0&0&\varrho_{14}\\
0&\varrho_{22}&\varrho_{23}&0\\
0&\varrho_{23}^\ast&\varrho_{33}&0\\
\varrho_{14}^\ast&0&0&\varrho_{44}
\end{array}\right),~~~~~\mbox{with}~~\sum^4_{j=1}\varrho_{jj}=1\,,
\end{equation}
allow us to span the whole physically-allowed regions of the planes studied in this work, boundaries included. We will thus use the states in Eq.~(\ref{Xstates}) as our starting ansatz to seek analytical candidates for extremality. A posteriori, this appears as a natural choice, as all known MEMS~\cite{munro} and states that maximize discord at fixed entanglement~\cite{james} fall into this class.

\begin{figure}[t]
\includegraphics[width=5.4cm]{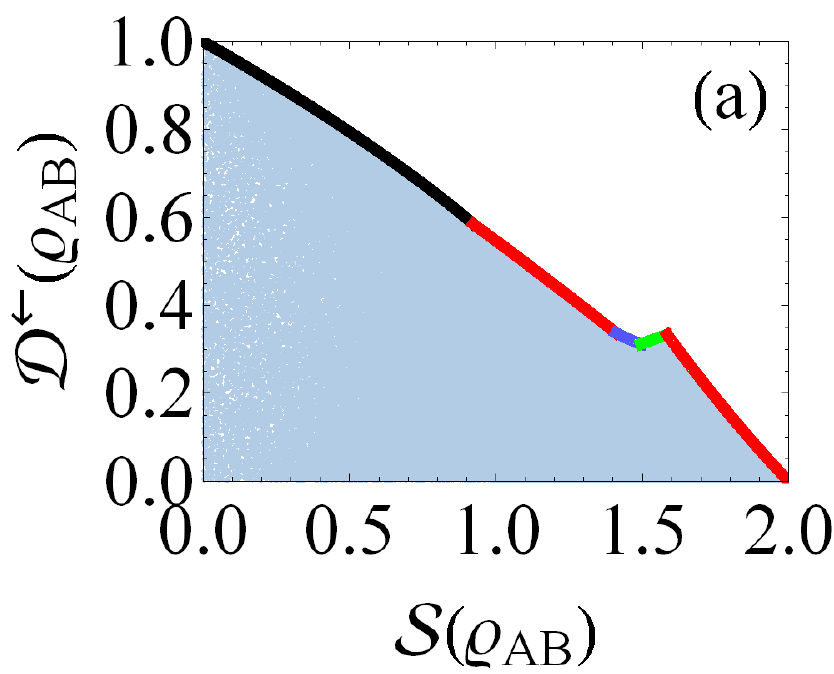}\hspace*{.1cm}
\includegraphics[width=5.4cm]{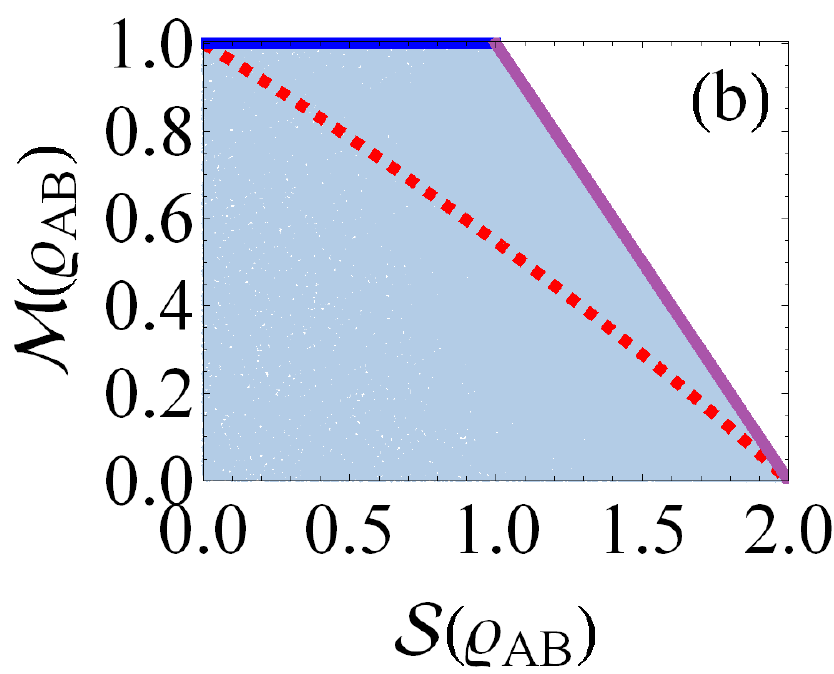}\hspace*{.1cm}
\includegraphics[width=5.4cm]{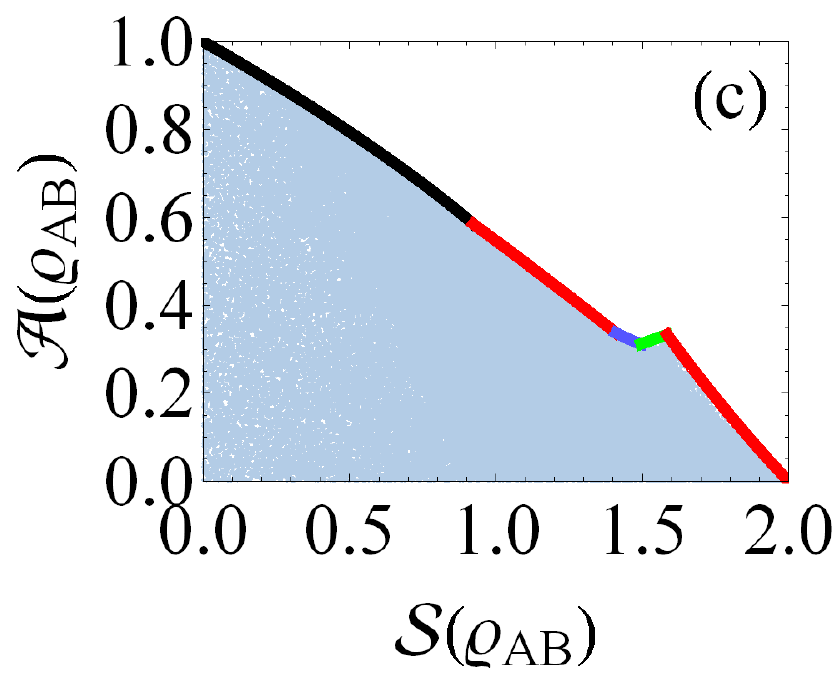}
\caption{(Color online). {{(a)}} Discord, {{(b)}} MID, and {{(c)}} AMID versus vNE  for $2\times 10^5$ random two-qubit states. The boundaries in {(a),(c)} correspond to the MQCMS of Table~\ref{azz}. In {(b)}, the extremal states are $\varrho_{AB}^\beta$ (horizontal, blue  segment) and $\varrho_{AB}^\delta$ (oblique, purple  segment), while Werner states $\varrho_{AB}^W$ lie on the dashed (red) curve.}\label{figrandom}
\end{figure}

\begin{table}[b]\centering\begin{tabular}{ccc}
 \hline
 $\varrho_{AB}$ & ${\cal S}(\varrho_{AB})$ & color\\
\hline
    $\varrho^R_{AB}$ with $0\le a \le \frac13$, $r=r^\star(a)$ & $[0, 0.9231)$ & black \\
    $\varrho^W_{AB}$ with $0\le f \le 1$  & $[0.9231, 1.410)$ & red \\
    $\varrho^P_{AB}$ with $b=0$  & $[1.410, 1.497)$ & blue \\
    $\varrho^P_{AB}$ with $0\le b \le 1$, $a=a^\star(b)$  & $[1.497, 1.585)$ & green \\
    $\varrho^W_{AB}$ with $-\frac13\le f \le 0$  & $[1.585, 2]$ & red \\ \hline
\end{tabular}
\caption{\label{azz} Maximally quantum-correlated mixed states of two qubits, corresponding  range of the von Neumann entropy, and color code for the curves in Fig.~\ref{figrandom}{(a),(c)};  the values of the parameters $a^\star$ and $r^\star$ are given in~\cite{trasc}.}
\end{table}

In Fig.~\ref{figrandom}{(a)} we plot the distribution of discord versus vNE for a sample of $2 \times 10^5$ random two-qubit states. Although a profile similar to the one for ${\cal D}^{\leftarrow}$-vs-linear-entropy is retrieved \cite{notejamesvne,james}, including the peculiar region
around  ${\cal S}{=}\log_2 3$ where discord increases at the expense of entanglement, we reveal interesting differences regarding the classes of extremal states drawing the boundary of the physically allowed area. These have been determined, in our investigation, by looking for the conditions to impose on $\varrho^{X}_{AB}$ so as to achieve the absolute maximum of ${\cal D}^{\leftarrow}$ at fixed  ${\cal S}$. Recalling that analytic expressions for the discord of $X$ states are available \cite{alber}, the problem can be efficiently solved using the Lagrange-multiplier method, in a way similar to Refs.~\cite{munro,maiorca}, and searching for the stationary points of the function ${\cal D}^\leftarrow(\varrho^X_{AB}){+}\lambda[{\cal S}(\varrho^X_{AB}){-}\tilde{\cal S}]$ with $\tilde{\cal S}\in[0,2]$ being an assigned value of vNE. The problem can be solved analytically and the resulting boundary states are presented in the following.
Let us define three subfamilies of states:
the rank-3 class $\varrho^R_{AB}$ encompassing the MEMS for the relative entropy of entanglement~\cite{munro},
the Werner states $\varrho^W_{AB}$, and a two-parameter family $\varrho^P_{AB}$ studied in~\cite{james}.
Their density matrices are as in~\eq{Xstates} with, respectively,
\begin{eqnarray}
\label{rhoRWP}
\varrho^R_{11}&=&\frac{1-a}{2}\,,\quad  \varrho^R_{22}=a\,,\quad\varrho^R_{14}=\frac{r}{2}\,,\quad\varrho^R_{33}=0\,, \nonumber\\
\varrho^W_{11}&=&
\frac{1+f}{4}\,,\quad  \varrho^W_{22}=\varrho^W_{33}=\frac{1-f}{4}\,,\quad\varrho^W_{14}=\frac{f}{2}\,, \\
\varrho^P_{11}&=&
\varrho^P_{14}=\frac{a}{2}\,,\quad  \varrho^P_{22}=\frac{1-a-b}{2}\,,\quad\varrho^P_{33}=\frac{1-a+b}{2}\,, \nonumber
\end{eqnarray}
and $\varrho^{R,W,P}_{23}{=}0$.
The maximally quantum-correlated mixed states (MQCMS) according to quantum discord are then reported in Table~\ref{azz}.
We observe an intricate  profile of extremally ``discordant'' two-qubit states at set vNE, far more structured than any instance of MEMS~\cite{munro}, thus showing the nontrivial relation between quantum correlations and global state mixedness.

 Let us now address a similar study when MID is used as a nonclassicality indicator. We aim at finding whether MID could be effective in providing a simpler yet meaningful picture of the behavior of quantum correlations in two-qubit mixed states. As shown in Fig.~\ref{figrandom}{(b)}, we find indeed that the physically allowed region in the ${\cal M}$-vs-${\cal S}$ plane simplifies to a trapezium, whose upper extremity is spanned by two different extremal families only. The states attaining the horizontal (blue online) boundary in Fig.~\ref{figrandom}{(b)} belong to the so-called $\beta$ family~\cite{james} $\varrho^\beta_{AB}=\beta\ket{\phi_+}\bra{\phi_+}+(1-\beta)\ket{\psi_+}\bra{\psi_+}$ with  $\ket{\phi_+}{=}(\ket{00}{+}\ket{11})/\sqrt{2}$, $\ket{\psi_+}{=}(\ket{01}{+}\ket{10})/\sqrt{2}$ and $\beta{\in}[0,1]$. They have maximal MID, ${\cal M}{=}1$, regardless of $\beta$ and thus of ${\cal S}$. On the other hand, by simple geometrical considerations, we see that $\varrho^\delta_{AB}{=}\delta\varrho^{\beta=1/2}_{AB}{+}(1{-}\delta){\mathbbm{1}}/4$ ($\delta\in[0,1]$), having ${\cal M}(\varrho^\delta_{AB})= [(1{-}\delta)\log_2(1{-}\delta){+}(1{+}\delta)\log_2(1{+}\delta)]/2{=}2{-}{\cal S}(\varrho^\delta_{AB})$, fill up the diagonal (purple online) side of the trapezium.
Quite strikingly, however, the boundary state $\varrho^{\beta=1/2}_{AB}$ and all  the $\varrho^{\delta}_{AB}$'s on the oblique edge of the trapezium have strictly ${\cal D}^\leftarrow{=}{\cal D}^\rightarrow{=}0$ and are thus  classical-classical:  MID clearly overestimates quantum correlations, failing to meet the essential faithfulness requirement.

\begin{figure}[t]
\includegraphics[width=7cm]{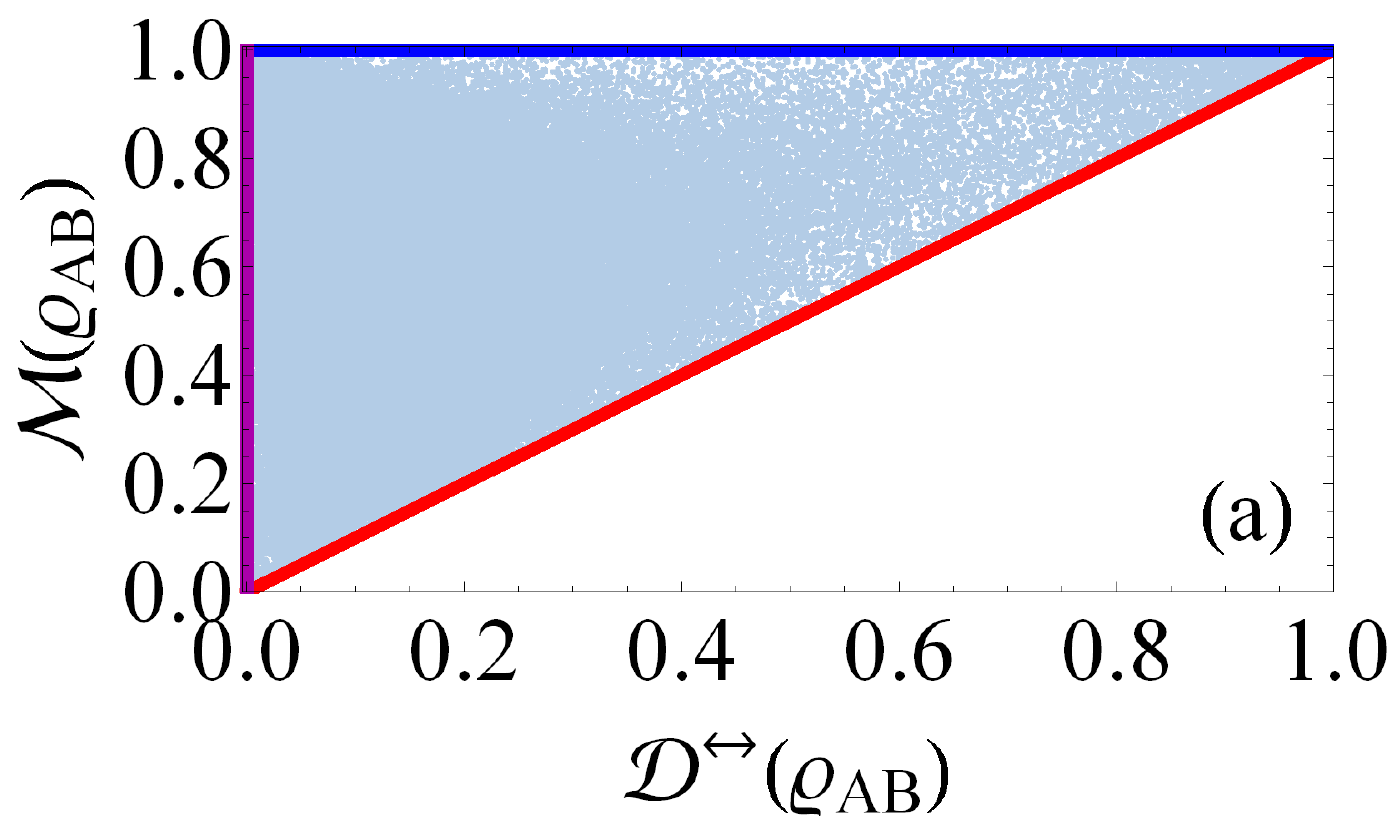}\hspace*{.1cm}
\includegraphics[width=7cm]{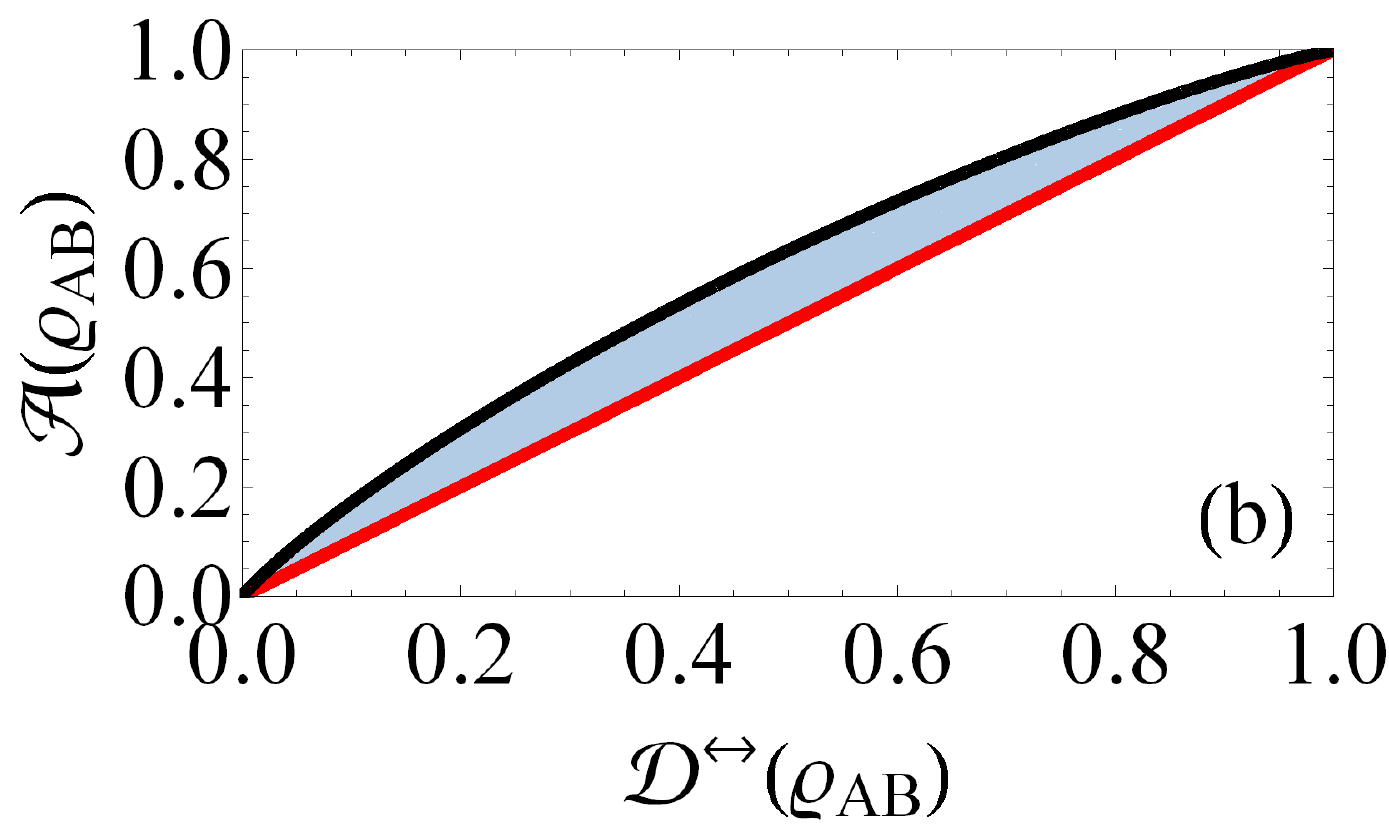}
\caption{(Color online). {(a)} MID and {(b)} AMID versus two-way discord ${\cal D}^\leftrightarrow{\equiv}\max[{\cal D}^\rightarrow,{\cal D}^\leftarrow]$, for $10^5$ random two-qubit states. The lower (red) boundary is spanned in both cases by pure and Werner states. The $\beta$ and $\delta$ families [see text] fill up the topmost horizontal (blue) and leftmost vertical (purple) boundaries of {(a)}. States  $\varrho^{\epsilon}_{AB}$ \cite{trasc2} sit on the upper (black) boundary in {(b)}.}
\label{figOmid}
\end{figure}

 To confirm this, we plot in Fig.~\ref{figOmid}{(a)} the distribution of MID versus discord for random two-qubit states. Here we consider the ``two-way'' discord ${\cal D}^\leftrightarrow{\equiv}\max[{\cal D}^\rightarrow,{\cal D}^\leftarrow]$, to ensure its vanishing only for classical-classical states  (while it can be ${\cal D}^\leftarrow(\varrho_{AB}){=}0$ but ${\cal D}^\rightarrow(\varrho_{AB})>0$ on classical-quantum states \cite{piani}, as discord is not strongly faithful). Perhaps not surprisingly, MID is in general a very loose upper bound to discord. The two indicators coincide on pure and Werner states, which span the lower boundary of the triangular region in Fig.~\ref{figOmid}{(a)}. As expected, $\varrho_{AB}^{\beta}$ yields the upper boundary at ${\cal M}{=}1$, while the vertical side at ${\cal D}^{\leftrightarrow}{=}0$ accommodates
classical-classical states (e.g.~$\varrho_{AB}^\delta$),  yet with nonzero MID up to its maximum value \cite{notemidstrana}.

\section{Ameliorated MID as a measure of nonclassical correlations}\label{secAMID}

The evident overestimation given by MID, which is a consequence of the `rigidity' of the projective bases used in its definition (as already pointed out in \cite{wu}) urges us to look for a more faithful figure of merit.
 The natural next step is to consider an {\it ameliorated} version of the measurement-induced disturbance (or AMID) where arbitrary complete projective measurements  can be performed, locally, on parties $A$ and $B$, and a subsequent optimization over any possible set of local projectors is achieved. We thus define AMID, for bipartite systems of any dimension, as
\begin{equation}
\label{mida}
{\cal A}(\varrho_{AB})={\inf}_{{\hat\Omega}_{AB}}[{\cal I}(\varrho_{AB})-{\cal I}(\varrho^{\hat{\Omega}}_{AB})]={\cal I}(\varrho_{AB})-{\cal I}_c(\varrho_{AB}),
\end{equation}
where ${\cal I}_c(\varrho_{AB}) \equiv {\sup}_{{\hat\Omega}_{AB}}{\cal I}(\varrho^{\hat{\Omega}}_{AB})$ and $\hat{\Omega}_{AB,kl}=\Omega_{A,k}\otimes\Omega_{B,l}$ is an arbitrary complete (bi-local) projective measurement over the composite system  (see also \cite{wu}). In particular, $\Omega_{A,k}$ and $\Omega_{B,l}$ are not necessarily made out of eigen-projectors and the search for the infimum over the set of $\hat{\Omega}_{AB}$'s entails the nontrivial optimization missing in MID \cite{MID}.
Our definition is further motivated by the earlier analyses discussed in~\cite{terhal}, where ${\cal I}_c$ in~\eq{mida} is recognized as the {\it classical mutual information} (optimized over projective measurements), a proper symmetric measure of classical correlations in bipartite states.
 AMID is thus  recast as the difference between total and classical mutual information, which has all the good prerequisites to be a {\it bona fide} measure of quantum correlations \cite{wu,pianietal}. From an operational point of view, the AMID can be interpreted as a measure of the ``infidelity'' of local broadcasting, a primitive task that can only be accomplished perfectly with classical-classical states \cite{piani}.

The evaluation of Eq.~(\ref{mida}) involves solving a nontrivial double-optimization problem. However, for two qubits  the techniques of Refs.~\cite{luo,alber} enable us to streamline the formal apparatus needed for the quantification of AMID. Any two-qubit state can be transformed by local unitaries (that leave AMID invariant by definition) into the form $\varrho'_{AB}{=}[{\mathbbm{1}}_{AB}{+}({\bf a}{\cdot}\hat{\bm\sigma}_A){\otimes}{\mathbbm{1}}_B{+}{\mathbbm{1}}_A{\otimes}({\bf b}{\cdot}\hat{\bm\sigma}_B)+\sum^3_{p=1}{\chi}_p(\hat{\sigma}_{A,p}{\otimes}\hat{\sigma}_{B,p})]/4$ with ${\bf a}$ and ${\bf b}$ the single-qubit Bloch vectors and ${\bm\chi}{=}\{\chi_1,\chi_2,\chi_3\}$ a correlation vector. Any projector $\hat\Omega_{j,k}$ for subsystem $j{=}A,B$, on the other hand, can be written as $\hat\Omega_{j,k}{=}U_j\hat\Pi_{j,k}U_j^\dag$ with $U_{j}{=}y_{j,0}{\mathbbm{1}}_j{+}i{\bf y}_{j}{\cdot}\hat{\bm\sigma}_{j}$ a unitary matrix such that $\sum^3_{p=0}y^2_{j,p}{=}1,~y_{j,p}{\in}[-1,1]$~\cite{luo,alber}. After some operator algebra, one has
${\varrho'}^{\hat{\Omega}}_{AB}=(\Omega_{A,k}\otimes\Omega_{B,l})\rho'_{AB}(\Omega_{A,k}\otimes\Omega_{B,l})=\Delta_{kl}(\Omega_{A,k}\otimes\Omega_{B,l})/4$,
where the vectors ${\bm\gamma}_{j}$ (depending solely on ${\bf y}_{j}$) are defined by  the relation $U^\dag_j\hat{\sigma}_{j,p}U_j{=}\alpha_{j,p}\hat\sigma_{j,1}{+}\beta_{j,p}\hat\sigma_{j,2}{+}\gamma_{j,p}\hat\sigma_{j,3}$, and $\Delta_{kl}{=}1{+}(-1)^k{\bf a}{\cdot}{\bm\gamma_A}{+}(-1)^l{\bf b}{\cdot}{\bm\gamma_B}{+}(-1)^{k+l}\sum^3_{p=1}\chi_{p}\gamma_{A,p}\gamma_{B,p}$~\cite{luo}. It is then convenient to introduce the new set of variables $\kappa_j{=}y^2_{j,0}{+}y^2_{j,3}$, $h_j{=}y_{j,0}y_{j,1}{+}y_{j,2}y_{j,3}$, $w_{j}{=}y_{j,1}y_{j,3}{-}y_{j,0}y_{j,2}$ and $l_j{=}1{-}\kappa_j$ and define $\mu(\kappa_A,h_A,w_A,\kappa_B,h_B,w_B){\equiv}{\cal I}(\varrho'_{AB}){-}{\cal I}({\varrho'}^{\hat{\Omega}}_{AB})$. By formulating and solving the conditions for stationarity of ${\cal A}$, as well as studying its behavior at the boundary of the parameter space, we finally find that there are at least two candidates for the global infimum in \eq{mida}. We thus obtain for AMID the following general upper bound
\begin{equation}\label{midasolved}
{\cal A} \le \min[\mu(1/2,0,1/2,1/2,0,1/2),\mu(1,0,0,1,0,0)]\,.
\end{equation}
For the relevant class of  $X$ states [\eq{Xstates}], \eq{midasolved} rigorously holds with equality, providing the exact analytical expression of their AMID [as well as of their classical mutual information ${\cal I}_c$, see \eq{mida}], and thus complementing the results of Refs.~\cite{luo,alber} on discord. 
 Crucially, the above analysis allows us to perform a complete  AMID-vs-vNE study  and investigate relations between MID,  AMID and  discord.

To begin with, we find quantitatively that \begin{equation}{\cal D}^\leftrightarrow \le {\cal A} \le {\cal M}\,.\end{equation} The rightmost bound is obvious by definition, whereas the leftmost one holds as ${\cal I}_c \le \{{\cal J}^\leftarrow,{\cal J}^\rightarrow\}$ \cite{terhal,wu}. This establishes an important {\it hierarchy} of quantumness indicators. While a plot of un-optimized-MID-vs-AMID  turns out to be totally analogous to the MID-vs-discord one [Fig.~\ref{figOmid}{(a)}], more interesting conclusions can be drawn from the exploration of the ${\cal A}$-vs-${\cal D}^\leftrightarrow$ plane, whose results are shown in Fig.~\ref{figOmid}{(b)}. First, Werner states (as well as pure states) still embody the lower-most boundary (where ${\cal M}{=}{\cal A}{=}{\cal D}^\leftrightarrow$). Second, as a consequence of the optimizations in its definition, AMID is a tighter upper bound to discord, occupying a much smaller area of the region above the Werner bound. Finally (and most importantly) there is no zero-discord state with nonzero AMID: the lower and upper boundaries to the physically meaningful region in the ${\cal A}$-vs-${\cal D}^\leftrightarrow$ plane are conjoined at the origin.

This provides a clearcut confirmation that AMID is a {\it strongly faithful} measure of genuinely quantum correlations, which vanishes if and only if a state is classical-classical \cite{piani}, is furthermore intrinsically symmetric, and can be then adopted as a rightfully valid and well motivated
 alternative to quantum discord.
The  states on the  upper boundary in the ${\cal A}$-vs-${\cal D}^\leftrightarrow$ plane can be  found again using the Lagrange multiplier method \cite{trasc2}.

We conclude by studying AMID vs vNE for arbitrary two-qubit random states. Most notably [see Fig.~\ref{figrandom}{(c)}], the physically allowed region in the $({\cal S},{\cal A})$ plane is found precisely congruent to the one in the $({\cal S},{\cal D}^\leftarrow)$ one [Fig.~\ref{figrandom}{(a)}] and admits {\it the same}  boundaries [Table \ref{azz}] (for those  states, ${\cal A}{=}{\cal D}^\leftarrow{=}{\cal D}^\rightarrow$). We can thus promote the interpretation of the set of states in Table~\ref{azz} as {\it universal} two-qubit MQCMS, being simultaneously extremal for  discord and AMID, at fixed vNE.  This  highlights a fascinating connection between such entropic nonclassicality indicators, that impose inequivalent orderings on partially nonclassical states, but yield identical prescriptions for extremality.

\section{Conclusions}\label{secC}
 We performed some significant steps towards the proper distinction between classical and quantum correlations in composite quantum systems. Our starting point was a self-contained investigation of the structure of quantum correlations in  two-qubit entropic spaces using  the most widespread ``measures'' of nonclassicality, namely discord \cite{OZ} and MID \cite{MID}. Going  beyond a mere hierarchical state classification, our analysis  naturally led to the proposal of adopting AMID, a general strongly faithful quantum correlation quantifier linked to minimal state-disturbance after optimized local measurements \cite{wu}, and amounting to the quantum counterpart of the classical mutual information \cite{terhal}. We explicitly computed  AMID on $X$ states and provided an exact upper bound (conjectured to be tight based on numerical evidence) on AMID for general two-qubit states, unveiling  interesting connections between AMID and discord, that include the sharing of the same set of extremal states (MQCMS) in the von Neumann entropic plane.

  The MQCMS can be rightfully regarded as the two-qubit states whose quantum correlations are maximally robust against state mixedness, and are thus set to play a key role in realistic (noisy) implementations of quantum information schemes based on nonclassicality of correlations as a resource \cite{piani,cinesi}, such as the ``deterministic quantum computation with one quantum bit'' \cite{DattaQC}. In this respect,   we remark that the engineering of MQCMS is feasible via both atom-light interfaces and all-optical setups~\cite{noiMegan,cinesi,mauroexp}, which adds an appealing feature of experimental demonstrability to our work.
  Our methods can be extended to faithfully investigate quantum correlations in  higher dimensional systems. Very recently, a comparative study of discord, MID, and  AMID has been reported for two-mode Gaussian states of continuous variable systems \cite{mista}.
We hope that our findings will motivate the search for  direct detection schemes for AMID in the context of quantum communication for general bipartite quantum systems, amenable of practical verification.

\bigskip

{\it Acknowledgments.} We thank V.\,P. Belavkin, S. Campbell, A. Datta, T. Paterek, M. Piani, F. Plastina, F.\,L. Semi\~ao, C. Theron, and A. Winter for valuable discussions on discord and MID. MP was supported by the UK EPSRC (EP/G004579/1).


\bigskip

\begin{thebibliography}{99}
\bibitem{piani}
Piani M, Horodecki P and Horodecki R 2008
{\it Phys. Rev. Lett.} \textbf{100} 090502

\bibitem{cinesi}  Xu J-S,  Xu X-Y,  Li C-F,  Zhang C-J,  Zou X-B and Guo G-C 2010 {\it Nat. Commun.} {\bf 1} 7

\bibitem{DattaQC}
Datta A, Shaji A and  Caves C M 2008 {\it Phys. Rev. Lett.} \textbf{100} 050502;
 Lanyon B P, Barbieri M, Almeida M P and  White A G 2008 {\it Phys. Rev. Lett.} {\bf 101} 200501

\bibitem{OZ} Ollivier H and Zurek W H 2001 {\it Phys. Rev. Lett.} {\bf 88} 017901

\bibitem{HV} Henderson L  and Vedral V 2001 {\it J. Phys. A} {\bf 34} 6899

\bibitem{terhal}  Terhal B M,  Horodecki H,  Leung D W and DiVincenzo D P 2002 {\it  J. Math. Phys.} {\bf 43} 4286;
DiVincenzo D P, Horodecki M, Leung D W,  Smolin J and  Terhal B M 2004 {\it  Phys. Rev. Lett.} {\bf 92} 067902

\bibitem{MID} Luo S 2008 {\it Phys. Rev. A} {\bf 77} 022301

\bibitem{wu} Wu S,  Poulsen U V and  M{\o}lmer K 2009 {\it Phys. Rev. A} {\bf 80} 032319

\bibitem{dissonance}
Modi K,  Paterek T,  Son W,  Vedral V and  Williamson M 2010 {\it Phys. Rev. Lett.} {\bf 104} 080501;  Daki\'c B,  Brukner C and  Vedral V 2010 {\it Phys. Rev. Lett.} {\bf 105} 190502

\bibitem{pianietal}
Piani M, Gharibian S, Adesso G, Calsamiglia J, Horodecki P and Winter A 2011  {\it Phys. Rev. Lett.} {\bf 106} 220403


\bibitem{varissimo} Datta A and  Gharibian S 2009 {\it Phys. Rev. A} {\bf 79}  042325; Datta A 2009 {\it Phys. Rev. A} {\bf 80} 052304;
 Ferraro A,  Aolita L,  Cavalcanti D,  Cucchietti F M and  Acin A 2010 {\it Phys. Rev. A} {\bf 81} 052318;
Mazzola L, Piilo J and  Maniscalco S 2010 {\it Phys. Rev. Lett.} {\bf 104} 200401;
 Adesso G and Datta A 2010 {\it Phys. Rev. Lett.} {\bf 105} 030501;
 Bradler K, Wilde M M,  Vinjanampathy S and Uskov D B 2010 {\it  Phys. Rev. A} {\bf 82} 062310.

\bibitem{operdiscord}  Cavalcanti D, Aolita L, Boixo S, Modi K, Piani M and Winter A 2011
{\it Phys. Rev. A} {\bf 83} 032324; Madhok V and  Datta A 2011 {\it Phys. Rev. A} {\bf 83} 032323


\bibitem{discordsymm} Maziero J,  Celeri L C and  Serra R 2010 arXiv:1004.2082
\bibitem{james}  Al-Qasimi A and James D F V  2011 {\it Phys. Rev. A} {\bf 83} 032101
\bibitem{maiorca} Galve F, Giorgi G L  and  Zambrini R 2011 {\it Phys. Rev. A}  {\bf 83} 012102
\bibitem{qbio}  Abbott D, Davies P C W, and   Pati A K (eds.), {\it Quantum Aspects of Life} (Imperial College Press, London, 2008)
\bibitem{qtech} Kimble H J 2008  {\it Nature} {\bf 453} 1023
\bibitem{horodecki} Horodecki R,  Horodecki P,  Horodecki M  and Horodecki K 2009 {\it Rev. Mod. Phys.} {\bf 81} 865
\bibitem{munro}  Wei T-C, Nemoto K, Goldbart P M,  Kwiat P G,  Munro W J and Verstraete F 2003 {\it Phys. Rev. A} {\bf 67} 022110
\bibitem{noiMegan} Adesso G, Campbell S, Illuminati F and Paternostro M 2010 {\it Phys. Rev. Lett.} {\bf 104} 240501
\bibitem{mauroexp} Chiuri A, Vallone G, Paternostro M and Mataloni P 2011 arXiv:1105.2109
\bibitem{luo} Luo S 2008 {\it Phys. Rev. A} {\bf 77} 042303
\bibitem{alber} Ali M,  Rau A R P  and  Alber G 2010 {\it  Phys. Rev. A} {\bf 81} 042105
\bibitem{notejamesvne}A similar study has been recently pursued for discord vs  linear entropy~\cite{james}. Here we use vNE as a more  ``compatible'' measure of mixedness, given the entropic nature of discord and MID.
\bibitem{trasc} In Table~\ref{azz}, $r^\star$ is the solution to $(1{-}\sqrt{a^2+r^2}) \log_2(1{-}\sqrt{a^2{+}r^2} )+(1{+}\sqrt{a^2{+}r^2}) \log_2(1{+}\sqrt{a^2+r^2}) = 2 a \log_2(a) + (1{-}a) \log_2(1{-}a)-(1{+}a) \log_2[(a+1)/{4}]$, while $a^\star$ solves $\log_2[\frac{ (b^2-1) (a^2+b^2-1)}{4(a-1)^2-4b^2}]+2 \sqrt{a^2+b^2} \tanh ^{-1} (\sqrt{a^2+b^2} )+(a+b) \log_2(1-a-b)+(a-b) \log_2(1-a+b)-2 a \log_2(2 a)+2 b \tanh ^{-1}b= -a/2$.

\bibitem{notemidstrana}
We remark that the definition of MID \cite{MID} is ambiguous on Bell diagonal states [that include the boundary states in Fig.~\ref{figrandom}(b)], whose marginal density matrices are maximally mixed states in any local basis, thus making the choice of the local eigenbases not unique (see also \cite{wu}). In particular, two Bell diagonal states can exist which are equivalent up to local unitary operations, and should thus contain the same amount of correlations, yet their MID may be zero or nonzero up to an arbitrary degree depending on the convention on the local eigenbases. In this respect, the boundaries in Fig.~\ref{figrandom}(b) should be understood as limiting cases of nearly-classical $X$ states with a slightly broken degeneracy in the marginal spectra.


\bibitem{trasc2}
The boundary states are $\varrho^{\epsilon}_{AB}{=}(1-\epsilon)\ket{\phi(p)_+}\bra{\phi(p)_+}+\epsilon\ket{01}\bra{01}$, with $\ket{\phi(p)_+}=\sqrt{p}\ket{00}+\sqrt{1-p}\ket{11}$, where $p \in [0.5,1]$ is the solution, at fixed $\epsilon\in[0,{\sim}0.2032]$, to the  equation $\mu(1/2,0,1/2,1/2,0,1/2){=}\mu(1,0,0,1,0,0)$.
Their discord is ${\cal D}^\leftrightarrow(\varrho^{\epsilon}_{AB}) = (1-2\epsilon)\log_2(\sqrt{\epsilon^{-1}-1}){-}\log_2(\sqrt{1-p})-(1-v_1)\log_2(1-v_1)-v_1\log_2(v_1)+
\sqrt{\frac{v_2}{4}}\log_2\left(\frac{1-\sqrt{v_2}}{1+\sqrt{v_2}}\right)$ and their AMID is ${\cal A}(\varrho^{\epsilon}_{AB})=(\epsilon{-}{v_1})\log_2(1{-}p){+}(1{-}{v_1}) \log_2\left(\frac{1-\epsilon}{1-{v_1}}\right)$,
where  $v_1{=}1-p(1-\epsilon)$,  $v_2{=}1+4\epsilon(\epsilon-v_1)$.
The states $\varrho^{\epsilon}_{AB}$ also enter in the maximization of discord at set classical correlations \cite{maiorca}.

\bibitem{mista}
Mista Jr L, Tatham R, Girolami D, Korolkova N and Adesso G 2011 {\it Phys. Rev. A} {\bf 83} 042325

\end{thebibliography}
\end{document}